\documentclass[aps,twocolumn,raggedbottom,prl,showpacs,nobalancelastpage,amssymb,groupedaddress]{revtex4}
\usepackage[latin1]{inputenc}
\usepackage[english]{babel}
\usepackage{graphicx}
\usepackage{psfrag}
\usepackage{amssymb}
\usepackage{amsmath}
\usepackage{amscd}
\usepackage{eucal}
\usepackage{color}
\usepackage{bm}
\usepackage{cancel}
\newcommand{\bi}{\bibitem}

\begin{document}

\title{Universal Features of Spin Transport and Breaking of Unitary Symmetries}
\author{
Ph.~Jacquod$^1$ and \.{I}.~Adagideli$^2$}
\affiliation{
$^1$Physics Department, University of Arizona, 
1118 E. 4$^{\rm th}$ Street, Tucson, AZ 85721, USA \\
$^2$Faculty of Engineering and Natural Sciences, Sabanci University, Orhanli-Tuzla, Istanbul, Turkey}

\date{\today}
\begin{abstract}
When time-reversal symmetry is broken,
quantum coherent systems with and without spin rotational symmetry
exhibit the same universal behavior in their electric transport properties.
We show that spin transport discriminates between these two cases. 
In systems with large charge conductance, spin transport is essentially insensitive
to the breaking of time-reversal symmetry. However, in the opposite limit of a 
single exit channel, spin currents vanish identically in the 
presence of time-reversal symmetry, but are turned on by breaking it with an orbital 
magnetic field.
\end{abstract}
\pacs{72.25.Dc,73.23.-b,75.76.+j}
\maketitle

{\bf Introduction.}
Fifty years ago, Dyson showed that ensembles of unitary matrices that are invariant under 
general symmetry groups reduce to the direct product of three irreducible
ensembles~\cite{Dys62}. These three circular ensembles
are labelled
by an index $\beta=1,2,4$ and are respectively
invariant under the transformations
\begin{subequations}
\begin{eqnarray}
S & \rightarrow & U^T S U \, ,  \, {\rm orthogonal \, ensemble}, \,  \beta=1 ,\\
S & \rightarrow & U S V \, ,  \, {\rm unitary \, ensemble}, \, \beta=2  , \\
S & \rightarrow & W^R S W \, , \,  {\rm symplectic \, ensemble}, \, \beta=4 , 
\end{eqnarray}
\end{subequations}
where $S$ is an element of the ensemble, $U$ and $V$ are arbitrary  unitary matrices, $W$ is a
quaternion~\cite{quat} unitary matrix, $U^T$ is the transpose of $U$ and
$W^R = \sigma^{(y)} W^T \sigma^{(y)}$ is the dual of $W$~\cite{Mehta}. Here and below, $\sigma^{(\mu)}$, $\mu=x,y,z$ is a Pauli matrix.
This classification carries over to electronic quantum transport~\cite{Bee97}, 
where the three classes
are defined by time-reversal symmetry (TRS), an antiunitary symmetry. 
Systems without TRS have a scattering matrix in the $\beta=2$ ensemble, 
while systems with TRS 
are differentiated by whether the TRS operator squares to $+1$ ($\beta=1$) or $-1$
($\beta=4$). When TRS is preserved, breaking spin rotational symmetry (SRS) induces a crossover $\beta = 1 \rightarrow 4$,
however when TRS is broken, breaking SRS only 
doubles the size of the scattering matrix as a Kramers degeneracy gets removed. This does
not generate a new ensemble~\cite{Dys62,Bee97,Ale01}.

Quantum corrections to electric transport depend on the symmetry index $\beta$, but
are independent of the size $N$ of the
scattering matrix (giving the total number of transport channels from and to the scatterer) 
for large $N$~\cite{Bee97}. According to the above classification, universality in 
charge transport is therefore mostly determined by
the antiunitary TRS. 
Recent investigations of spin transport showed that the magnetoelectric spin conductance
\begin{equation}\label{eq:stcoeff}
\mathcal{T}^{(\mu)}_{ij} =  {\rm Tr}[S_{ij}^{\dagger} \sigma^{(\mu)} S_{ij}] \, , 
\end{equation}
constructed from the 
transmission block $S_{ij}$ of the scattering matrix connecting terminals $i$ and $j$,
also exhibits a character of universality~\cite{Bar07,Naz07,Kri08,Bar09,Kri09} in that 
${\rm var} \, \mathcal{T}^{(\mu)}_{ij} = 4 N_{i} (N_{i} -1) N_{j} /N (2 N-1) (2 N-3)$ for $\beta=4$.
Here, $N_{i,j}$ gives the number of transport channels between the system and terminals $i,j$, and 
$N=\sum_i N_i$. 
The spin conductance fluctuates about zero average, $\langle \mathcal{T}^{(\mu)}_{ij} \rangle  = 0$ 
and the resulting, typically nonzero spin current is
generated by the presence of a SRS breaking field. In the $\beta=4$ ensemble
one usually takes the latter field as spin-orbit interaction (SOI). 
In the absence of SOI, one
has $\mathcal{T}^{(\mu)}_{ij} \equiv 0$. This is the case for $\beta=1$ and, if Dyson's three-fold way applies
to spin transport, for $\beta=2$. In this manuscript we demonstrate that spin transport
discriminates between systems with and without SRS even when TRS is broken. Accordingly, a novel
kind of universality emerges in systems with broken SRS and TRS, with
charge transport properties given by those of the $\beta=2$ ensemble,
but with specific spin transport properties.
The latter are similar
to those of the $\beta=4$ ensemble at large $N$, a finding already reported in Ref.~\cite{Caio}
for specific four-terminal setups, 
but deviate from it at small $N$.
Our finding does not invalidate Dyson's classification-- the latter gives a complete 
classification of unitary scattering matrices and unless one introduces chiral or particle-hole 
symmetries~\cite{ver93,alt97},
there is no new ensemble to be found. Instead our  point is that spin-dependent observables
define two sub-ensembles of the $\beta=2$ ensemble, depending on whether they commute or not with
the scattering matrix.  In other words, we find that while 
universality in charge
transport is affected only by the antiunitary TRS
universality in spin transport depends on both antiunitary (TRS) and unitary (SRS) symmetries.

{\bf The model.} 
We consider a mesoscopic conductor connected to any number of external electron reservoirs.
There is no ferromagnetic exchange anywhere
in the system, nor is there spin accumulation in the reservoirs, thus injected currents are not polarized. 
We neglect spin relaxation in the terminals. The magnetoelectrically
generated spin current due to the presence of SOI inside the cavity
is determined by the spin-dependent transmission coefficients of Eq.~(\ref{eq:stcoeff}).
For instance, in the simple case of a two-terminal setup,  the generated spin current in the right lead
along the polarization axis
$\mu = x,y,z$ is given by
\begin{eqnarray}\label{eq:twoterms}
I_{\rm R} ^{(\mu)} & = & (e^2 V/h) {\cal T}_{\rm RL}^{(\mu)} \, ,
\end{eqnarray}
with the voltage bias $V$ applied across the sample.

{\bf Semiclassical calculation.}
We first calculate the average and mesoscopic fluctuations of the
spin transmission coefficients using the semiclassical theory of transport~\cite{Ric02,Bro06},
extended to take spin transport into account~\cite{Wal07,Ada10}. We write
(See Supplemental Material~\cite{suppl})
\begin{equation}\label{tr_prob_semicl}
{\mathcal T}_{ij}^{(\mu)} =  \!\! \int_i \! {\rm d} y \!\int_j \! {\rm d} y_0
\sum_{\gamma,\gamma'}
A_\gamma A_{\gamma'}^* e^{i(S_\gamma-S_{\gamma'})}
{\rm Tr}[U_\gamma \sigma^{(\mu)} U_{\gamma'}^\dagger] \, .
\end{equation}
The sums run over all  trajectories starting at $y_0$ on a cross-section of
the injection lead $j$ and ending at $y$ on the exit lead $i$. Trajectories have a stability
given by $A_\gamma$, which includes a prefactor $(2 \pi i \hbar)^{-1/2}$ as well as a
Maslov index~\cite{maslov}, and $S_\gamma$ gives the classical action accumulated on $\gamma$, in units
of $\hbar$. SOI is incorporated in the matrices $U_\gamma$.  
The average spin conductance has been calculated semiclassically in Ref.~\cite{Ada10}.
In the absence of SOI, spins do not rotate, $U_\gamma = \sigma^{(0)}$ is the identity matrix, 
and one trivially obtains
${\mathcal T}_{ij}^{(\mu)} \equiv 0$.
The leading-order approximation is to consider $U_\gamma \in$
SU(2), where SOI
rotate the spin of the electron along unperturbed classical
trajectories~\cite{Wal07,Mat92}. 
In this manuscript, we will use this approximation because, even though it
neglects the geometric correlations reported in Ref.~\cite{Ada10}, it is appropriate for
our search of universality. At that level, 
the average spin conductance vanishes, 
$\langle {\mathcal T}_{ij}^{(\mu)} \rangle_{\rm semicl} = 0$~\cite{Ada10}, which agrees with the
random matrix theory (RMT) result of Ref.~\cite{Bar07}.

\begin{figure}
 \centering
 \includegraphics[width=0.99\columnwidth]{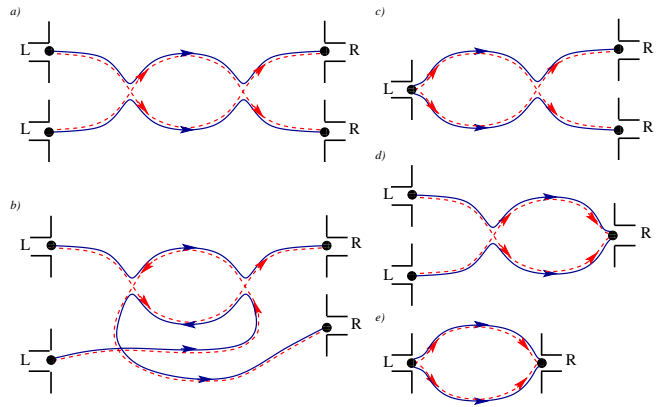}
\caption{\label{fig:semicl_ucf} (Color online) Semiclassical 
diagrams determining the conductance and spin conductance fluctuations
to leading order in the number $N \gg 1$ of transport channels. Blue (dark) and red
(light) trajectories travel in opposite direction in  diagram $b)$, which consequently vanishes in the
presence of a large magnetic flux piercing the loop. All other diagrams are insensitive to the breaking
of time-reversal symmetry.\\[-7mm]}
\end{figure}

Having established that 
the average spin conductance vanishes regardless of the presence or absence of TRS and SRS,
we next calculate spin conductance fluctuations. The leading-order diagrams contributing to 
${\rm var} [ {\mathcal T}_{\rm RL}^{\mu0}]_{\rm semicl} $ are shown in Fig.~\ref{fig:semicl_ucf}. They are the same
as those contributing to the (charge) transmission fluctuations [substituting $\sigma^{(\mu)}
\rightarrow \sigma^{(0)}$ in Eq.~(\ref{eq:stcoeff})]. In this case, Ref.~\cite{Bro06} found that the sum of
contributions $c)$, $d)$ and $e)$ cancel out, furthermore, contribution $b)$ vanishes
upon breaking of TRS. This can be achieved via a magnetic flux piercing the diagram's loop.
From Fig.~\ref{fig:semicl_ucf}, we see that contribution $b)$ is the only one that is flux-sensitive, because
the blue (dark) and the red (light) trajectories accumulate the same flux-phase. 
From a semiclassical point of view, this is the origin of the halving of the universal conductance 
fluctuations upon TRS breaking~\cite{Bee97}. Extending this calculation to ${\rm var} [{\mathcal T}_{ij}^{\mu}]_{\rm semicl}$,
we obtain that contributions $a)$, $b)$ and $c)$ are multiplied by a spin-dependent term
${\rm Tr}[U_{\gamma_5}^\dagger U_{\gamma_3}^\dagger \sigma^{(\mu)} U_{\gamma_3} 
U_{\gamma_2}] \times {\rm Tr}[U_{\gamma_2}^\dagger U_{\gamma_6}^\dagger \sigma^{(\mu)} U_{\gamma_6} 
U_{\gamma_5}] $, while contributions $d)$ and $e)$ are multiplied by
$|{\rm Tr}[U_{\gamma_5}^\dagger \sigma^{(\mu)} U_{\gamma_2}]|^2$ (See Supplemental Information for the
labelling of trajectory segments~\cite{suppl}). All these terms vanish in the
absence of SOI. In the presence of SOI, we evaluate them by
averaging over a uniform distribution of all $U_\gamma$'s over the SU(2) group, corresponding to totally
broken SRS.
Following the standard procedure of
performing orbital averages and spin averages separately, we obtain that, when SRS is totally
broken, 
contributions $a)$, $b)$ and $c)$ acquire a prefactor ($\langle ... \rangle_{\rm SU(2)}$ indicates an 
homogeneous average over the SU(2) group)
\begin{equation}
\langle {\rm Tr}[U_{\gamma_5}^\dagger U_{\gamma_3}^\dagger \sigma^{(\mu)} U_{\gamma_3} 
U_{\gamma_2}] \, {\rm Tr}[U_{\gamma_2}^\dagger U_{\gamma_6}^\dagger \sigma^{(\mu)} U_{\gamma_6} 
U_{\gamma_5}] \rangle_{\rm SU(2)} = 0 \, ,
\end{equation}
and thus vanish identically, 
while contributions $d)$ and $e)$ are multiplied by
\begin{eqnarray}\label{eq:semicl_vargmu}
\langle |{\rm Tr}[U_{\gamma_5}^\dagger \sigma^{(\mu)} U_{\gamma_2}]|^2 \rangle_{\rm SU(2)} = 1 \, .
\end{eqnarray}
We conclude that the semiclassical contributions to the spin conductance fluctuations 
are those with a correlated encounter at the exit terminal, which in particular
has the consequence that they are not sensitive to the breaking of TRS.

We obtain the variance of the spin conductance coefficients
as the sum of contributions $d)$ and $e)$, i.e.
\begin{equation}\label{eq:var_semicl}
{\rm var} [{\mathcal T}_{ij}^{(\mu)}]_{\rm semicl} =  (N_{i} N_{j} N - N_{i} N_{j}^2) \big/
N^3  \, .
\end{equation}
The key point is that this result holds both in the absence and in the presence of 
TRS, because both relevant contributions $d)$ and $e)$ are
sensitive neither to magnetic fluxes piercing their loops, nor to orbital magnetic field effects that do
not alter the ergodicity of the classical trajectories. Thus, Eq.~(\ref{eq:var_semicl}) gives the
leading-order semiclassical expression for the conductance variance, for systems without SRS (with SOI) 
in both cases of conserved or broken TRS, as well as in the intermediate regime of
partially broken TRS. Therefore, to leading order in the number $N \gg 1$ of transport chanels,
spin conductance fluctuations are insensitive to the breaking of TRS. In the next section, 
this result is confirmed using RMT.

{\bf Random matrix theory calculation.}
We next use the method of Ref.~\cite{Bro96} to calculate the RMT average and fluctuations of the spin
conductance. We write~\cite{Bar07}
\begin{subequations}
\begin{align}
  \label{eq:TfullTr}
  \mathcal{T}_{ij}^{(\mu)} &= \text{Tr}\,[Q_{\rm i}^{(\mu)}SQ_{\rm j}^{(0)}S^\dagger], \\
  [Q_{\rm i}^{(\mu)}]_{m\eta,n\nu} &= 
  \begin{cases}
    \delta_{mn}~\sigma^{(\mu)}_{\eta\nu}, & m \in i\\
    0, & \text{otherwise} \, ,
  \end{cases} \\
  [Q_{\rm j}^{(\mu)}]_{m\eta,n\nu} &= 
  \begin{cases}
    \delta_{mn}~\sigma^{(\mu)}_{\eta\nu}, & m \in j, \\
    0, & \text{otherwise} \, ,
  \end{cases}
\end{align}
\end{subequations}
where  $m$ and $n$ are channel indices,
$\eta$  and $\nu$ are spin indices and $\sigma^{(0)}$ is the $2 \times 2$
identity matrix. The trace in Eq.~(\ref{eq:TfullTr}) 
is taken over both sets of indices. We find that the average of the spin transmission vanishes
in all cases,
\begin{equation}\label{eq:sping_avg}
\langle {\mathcal T}_{ij}^{(\mu)}  \rangle_{\rm RMT} = 0 \, .
\end{equation}
For the $\beta=4$ ensemble, this result was first obtained in Ref.~\cite{Bar07}. 
We further obtain 
\begin{subequations}\label{eq:var_RMT}
\begin{eqnarray} 
{\rm var} [{\mathcal T}_{ij}^{(\mu)}]_{\beta=2; \text{SRS}} &=& 0 \; , \\
{\rm var} [{\mathcal T}_{ij}^{(\mu)}]_{\beta=2; \xcancel{\rm SRS}} &=& 4  \frac{N_{i} N_{j} N - N_{i} N_{j}^2}
{N (4 N^2-1)} \; ,  \label{eq:var_RMT2} \\ 
{\rm var} [{\mathcal T}_{ij}^{(\mu)}]_{\beta=4} &=& 4 
\frac{N_{i} N_{j} (N-1)- N_{i} N_{j}^2}{N (2N-1) (2N-3)}
\; . \qquad \label{eq:var_RMT3}
\end{eqnarray}
\end{subequations}
Eq.~(\ref{eq:var_RMT3}) first appeared in Ref.~\cite{Bar07}, and expressions similar to Eq.~(\ref{eq:var_RMT2})
appeared in Refs.~\cite{Kri09,Caio} for two-terminal geometries.
We see that Eqs.~(\ref{eq:var_semicl}), 
(\ref{eq:var_RMT2}) and (\ref{eq:var_RMT3}) all agree in the limit $N_{i,j} \gg 1$, however,
while the semiclassical expression Eq.~(\ref{eq:var_semicl}) 
is valid only in that limit, Eqs.~(\ref{eq:var_RMT}) are exact
for any number of channels. Most interestingly, for a two-terminal setup with 
$N_{i}=1$, Eq.~(\ref{eq:var_RMT3}) gives 
${\rm var} [{\mathcal T}_{ij}^{(\mu)}]_{\beta=4} =0$. Together with Eq.~(\ref{eq:sping_avg})
this gives an identically vanishing spin conductance, in agreement with Ref.~\cite{Zha05}.
This restriction no longer applies once TRS is broken, as reflected in 
Eq.~(\ref{eq:var_RMT2}) -- breaking TRS can turn spin currents in two-terminal geometries, when the 
exit terminal carries a single transport channel.

{\bf Numerical simulations.}
We numerically confirm our findings
using the quantum mechanical spin kicked rotator model~\cite{Bar05}. It is represented by a
$2M \times 2M$ Floquet matrix~\cite{Bar05,Two03,Izr90} (See Supplemental Material~\cite{suppl})
\begin{subequations}
\label{eq:floquet}
\begin{align}
\mathcal{F}_{ll'}&=(\Pi U X U^\dagger \Pi)_{ll'}, \quad l,l' = 0,1,\ldots, M-1,\\
\Pi_{ll'}&=\delta_{ll'}e^{-i\pi (l+l_0)^2/M }\sigma_0, \\
U_{ll'}&=M^{-1/2}e^{-i2\pi ll'/M}\sigma_0,\\
X_{ll'}&= \delta_{ll'}e^{-i(M/4\pi)V(2\pi l/M)} \, .
\end{align}
\end{subequations}
The matrix $\Pi$ represents free ballistic motion, periodically interrupted by 
spin-independent and spin-dependent kicks given by the 
matrix $X$, and corresponding to scattering at the boundaries of the quantum dot, as well as
SOI. We choose
\begin{equation}
V(p) =  K\cos (p+\theta) \,\sigma_0 + K_\text{so}(\sigma_x\sin2p
+ \sigma_z\sin p) \, .
\end{equation}
The corresponding classical map is chaotic for kicking strength $K \gtrsim 7.5$, accordingly 
in nour search for universal behavior, we restrict
ourselves to that regime. The SO coupling strength 
$K_\text{so}$ is related to the SO rotation
time $\tau_\text{so}$ (in units of the stroboscopic period)
through
$\tau_\text{so} = 32\pi^2/K_\text{so}^2M^2$~\cite{Bar05}.
From (\ref{eq:floquet}), we construct the
quasienergy-dependent scattering matrix as
\begin{figure}
 \centering
 \includegraphics[width=0.85\columnwidth]{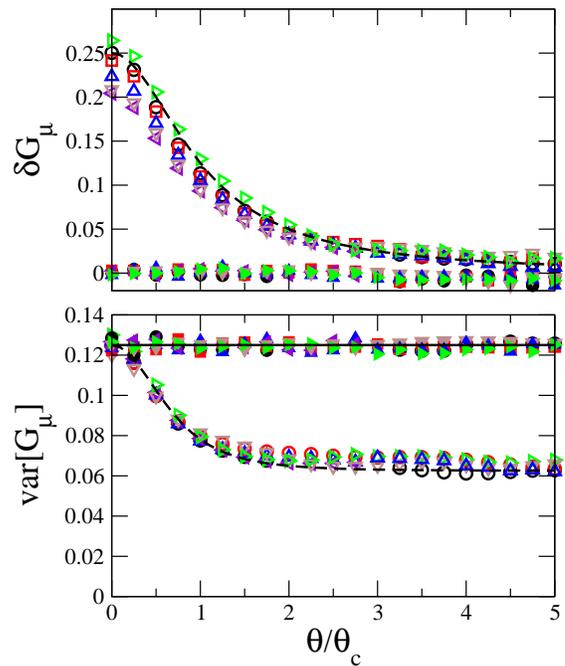}
\caption{\label{fig:num_univ} (Color online) Weak localization corrections to (top), and 
variance of (bottom) the charge (empty symbols) and spin (full symbols) conductance for the 
two-terminal quantum kicked 
rotator of Eqs.~(\ref{eq:floquet}). Parameters are
$\tau_{\rm D}=10, 20$, $K=40, 60, 80, 90$, $K_{\rm so} = 120 \, K_{\rm soc}$
and $M=128, 256, 512$. The dashed lines indicate the RMT predicted crossover from $\beta=4$
to $\beta=2$~\cite{Bar05}. Our semiclassical prediction of Eq.~(\ref{eq:var_semicl}) is 
illustrated by the straight black line in the bottom panel. For all data, $N > 10$.\\[-7mm]
}
\end{figure}
\begin{equation}
S(\varepsilon)=P[e^{-i\varepsilon}-{\mathcal F}(1-P^TP)]^{-1}{\mathcal F}P^T,
\end{equation}
with $P$ a $2N \times 2M$ projection matrix
\begin{equation}
 P_{k\alpha,k'\beta} =
\begin{cases}
    \delta_{\alpha\beta} &\text{if } k' = l^{(k)}, \\
    0 &\text{otherwise}.
\end{cases}
\end{equation}
The $l^{(k)}$ ($k=1,2,\ldots, 2N $, labels the modes) give
the position in phase space of the attached leads. The mean dwell
time $\tau_\text{D}$ is given by
$\tau_\text{D} = M/N$. The parameter $K_\text{so}$ breaks SRS over a scale 
$K_\text{soc} = 4 \pi \sqrt{2}/M \tau_\text{D}^{1/2}$ corresponding to 
$\tau_{\rm so} = \tau_{\rm D}$,  and 
$\theta$ breaks time-reversal symmetry over
a scale $\theta_c = 4 \pi/K M \tau_\text{D}^{1/2} $ when $l_0$ is finite~\cite{Bar05}.
In our numerics we fix $l_0=0.14$. 
When $K \gg 1$ and $\theta/\theta_c \gg 1$, the charge conductance properties are those of the 
$\beta=2$ ensemble, while for $\theta=0$ and $K_{\rm so}/K_{\rm soc} \gg 1$ they are those of the 
$\beta=4$ ensemble~\cite{Bar05}. In our numerics, we fix $K_{\rm so}/K_{\rm soc} = 120$ and
vary $\theta$ to gradually break TRS, starting from $\theta=0$.
For simplicity, we specify to two-terminal setups and accordingly calculate the dimensionless
spin conductance defined by Eq.~(\ref{eq:twoterms}) as $G_\mu = \mathcal{T}_{\rm RL}^{(\mu)}$
for $\mu=z$. We checked, but do not show, that numerical results remain the same if instead we
consider $\mu=x,y$.
\begin{figure}
 \centering
 \includegraphics[width=0.875\columnwidth]{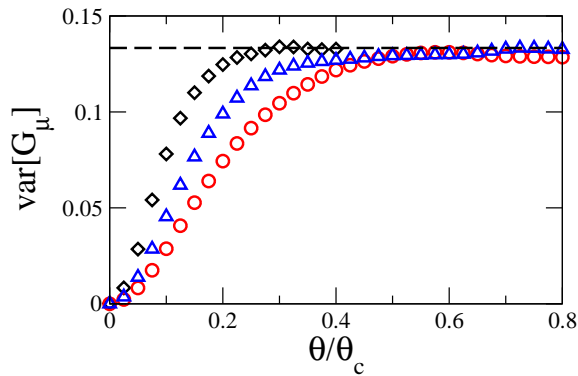}
\caption{\label{fig:num_N1} Spin conductance fluctuations for the quantum kicked rotator with SOI
defined in Eq.~(\ref{eq:floquet}) vs. the rescaled TRS breaking parameter $\theta/\theta_c$
for $N_{\rm R}=N_{\rm L}=1$. For $\theta=0$, one is in the $\beta=4$ ensemble and TRS 
forces the spin conductance to vanish~\cite{Zha05}. Breaking TRS results in a finite
variance of the spin conductance.
Dashed line: RMT prediction ${\rm var}[G_\mu] = 4/30$ for $N_{\rm R}=N_{\rm L}=1$
[see Eq.~(\ref{eq:var_RMT2})]. Data correspond to $K=45$, $K_{\rm so} = 120 \, K_{\rm soc}$, 
with $M=128$ (red circles), 256 (blue triangles) and 512 (black diamonds). The curves do not lie on top of one
another, because the rescaling of the horizontal axis with $\theta_c$ assumes 
$N_{\rm R, L}\gg1$~\cite{Bar05}.\\[-7mm]}
\end{figure}

Fig.~\ref{fig:num_univ} first shows data for quantum corrections to the charge and spin conductance,
as TRS is gradually broken. The top panel shows that weak localization corrections to the 
charge conductance are damped by a Lorentzian $\sim [1+(\theta/\theta_c)^2]^{-1}$ as predicted
by RMT~\cite{Bee97} and semiclassics~\cite{Ric02}. There is no weak localization correction to the average 
spin conductance, both with and without TRS, in agreement with Ref.~\cite{Bar07}.
The bottom panel shows that charge conductance fluctuations are halved upon TRS breaking 
and their behavior  agrees well with theoretical predictions. The situation is entirely
different, however, for the spin conductance fluctuations, which are essentially insensitive to the 
breaking of TRS. This is in agreement with our predictions, Eqs.~(\ref{eq:var_semicl}) 
and (\ref{eq:var_RMT}) for the large number of channels $N>10$ 
considered in all data in Fig.~\ref{fig:num_univ}. The new universal behavior corresponding to 
broken SRS and TRS emerges at larger $\theta$, where the charge conductance corresponds to the
$\beta=2$ Dyson ensemble, while the spin conductance is essentially the same as that of the $\beta=4$ ensemble.

Fig.~\ref{fig:num_N1} best illustrates the new universal behavior. When the
exit lead carries a single transport channel, TRS requires that the spin conductance vanishes~\cite{Zha05},
regardless of the presence or absence of SRS. 
Fig.~\ref{fig:num_N1} shows that, when SRS is broken, breaking TRS turns spin currents on, whose
variance is given by Eq~(\ref{eq:var_RMT2}) once TRS is totally broken. Note that the magnitude of the
field necessary to break TRS for $N_{\rm R, L}=1$ becomes smaller and smaller in the semiclassical limit,
$M \rightarrow \infty$ as the dwell time grows in that limit, $\tau_{\rm D} \sim M$.

{\bf Conclusions.}
By direct calculation we have shown that the spin conductance is an observable that is sensitive
to the presence or absence of SRS even when TRS is broken. Breaking of SRS is necessary
to magnetoelectrically generate a spin current, thus to acquire a finite spin conductance, but
the latter is affected by TRS only when there are very few transport channels.
Accordingly, we conclude that the $\beta=2$
universality class splits into two different subsets
for spin transport. In both cases, charge transport properties correspond
to the $\beta=2$ class, however, the spin conductance vanishes identically when SRS is preserved, 
but exhibits a universal behavior when it is broken, see Eq.~(\ref{eq:var_RMT2}). 
Spin and charge transport
universality classes are related to TRS and SRS in Table.~\ref{table1}.
Examples of systems with broken SRS and TRS include 
spin-orbit coupled systems under not too strong external magnetic fields,
systems with spin textures and even spin valves with non-aligned magnetizations. 
Breaking TRS without 
breaking SRS is possible in systems with orbital magnetic field effects stronger than Zeeman effects, such as
few-channel n-doped GaAs quantum dots in fields of the order of few tens of milliTeslas~\cite{Bee97}.

\begin{table}[t]
\begin{tabular}{| c | c || c | c | } 
\hline
{\bf TRS} & {\bf SRS} & {\bf Charge transport} & {\bf Spin transport} \\
\hline
Yes & Yes & $\beta=1$ & $\beta=1$; $G_\mu \equiv 0$ \\
\hline
Yes & No & $\beta=4$ & $\beta=4$; Eqs.~(\ref{eq:sping_avg}) and (\ref{eq:var_RMT3}) \\
\hline
No & Yes & $\beta=2$ & $G_\mu \equiv 0$ \\
\hline
No & No & $\beta=2$ &   Eqs.~(\ref{eq:sping_avg}) and (\ref{eq:var_RMT2}) \\
\hline
\end{tabular}
\caption{\label{table1} Universality behavior of charge and spin transport properties in the four
possible cases of broken or unbroken SRS and TRS. When both symmetries are broken, 
the spin transport properties correspond to those of the $\beta=4$ Dyson ensemble in the
limit $N_{\rm R},N_{\rm L} \gg 1$. Deviations from $\beta=4$
are given in Eq.~(\ref{eq:var_RMT}) for the spin conductance variance. They are largest 
for small number of channels.\\[-7mm]
}
\end{table}

We thank M. B\"uttiker for several interesting discussions at various stages of this project.

\begin{widetext}

\newpage

\renewcommand{\thefigure}{S\arabic{figure}}
\renewcommand{\theequation}{S\arabic{equation}}
\setcounter{equation}{0}
\setcounter{figure}{0}

\begin{center}
 {\large \bf Supplemental Material}
\end{center}

{\bf Semiclassical approach to spin transport.}

The short-wavelength semiclassical approach to transport has been pioneered by
Stone and collaborators, and further developed to include quantum corrections
by Richter and Sieber~\cite{Ric02}. It is based on the scattering approach to transport,
where transmission amplitudes are replaced with their semiclassical expression
\begin{equation}\label{eq:t}
t_{ij} = \int_i {\rm d}y \int_j {\rm d}y_0 \sum_\gamma A_\gamma \exp[i S_\gamma] \, .
\end{equation}
The sums run over all  trajectories starting at point $y_0$ located at the cross-section of
the injection lead $j$ and ending at point $y$ at the cross-section of the exit lead $i$. 
The stability of the trajectory $\gamma$ is
given by $A_\gamma$, 
which includes a prefactor $(2 \pi i \hbar)^{-1/2}$ as well as a
Maslov index~\cite{maslov},
and $S_\gamma$ is the classical action accumulated on $\gamma$, in units
of $\hbar$. Charge conductances in units of twice the conductance quantum $2e^2/h$
are given by the transmission probability $|t_{ij}|^2$ which contains a double sum
over trajectories and four spatial integrals. In the semiclassical, short-wavelength limit, these
integrals reduce to two integrals~\cite{Ric02}, and one has
\begin{equation}\label{eq:T}
T_{ij} = |t_{ij}|^2 = \int_i {\rm d}y \int_j {\rm d}y_0 \sum_{\gamma,\gamma'}
\, A_\gamma A_{\gamma'} \exp[i (S_\gamma-S_{\gamma'})] \, .
\end{equation}
Noting that the stability is much less energy-dependent than $S_\gamma$, the integrals in the above expression for $\langle T_{ij} \rangle$
(averaged over a small, but finite energy interval) are evaluated 
via a stationary phase approximation which 
results in specific pairings of the trajectories
$\gamma$ and $\gamma'$~\cite{Ric02}. For the conductance fluctuations, one obtains 
\begin{equation}\label{eq:T2}
{\rm var} T_{ij} = \left\langle \left(\int_i {\rm d}y \int_j {\rm d}y_0 \sum_{\gamma,\gamma'}
\, A_\gamma A_{\gamma'} \exp[i (S_\gamma-S_{\gamma'})] \right)^2 \right\rangle  - \langle T_{ij} \rangle^2 \, .
\end{equation}
After a stationary phase approximation, this expression 
requires the pairing of four trajectories. The terms corresponding to disconnected pairings are cancelled
by $-\langle T_{ij} \rangle^2$. One is left with the five contributions shown in Fig.1 of the main
text. They were calculated in Ref.~\cite{Bro06}, which furthermore showed that the sum of contributions
c), d) and e) vanish. Thus only contributions a) and b) matter for the charge conductance.
Contribution b) vanishes when time-reversal symmetry is broken, thus the variance of the
conductance is divided by two.

The presence of spin-orbit interaction forces one to include spin rotation into the semiclassical
propagator of Eq.~(\ref{eq:t}). In the weak spin-orbit coupling limit one usually makes the
approximation in which the sole effect of the spin-orbit field is to rotate the spin along the 
unchanged 
classical trajectories. Mathur and Stone therefore  replaced Eq.~(\ref{eq:t}) by
\begin{equation}\label{eq:tU}
t_{i\sigma,j\sigma'} = \int_i {\rm d}y \int_j {\rm d}y_0 \sum_\gamma A_\gamma \exp[i S_\gamma]
(U_\gamma)_{\sigma,\sigma'} \, ,
\end{equation}
with $U_\gamma \in SU(2)$ encoding the spin rotation.
The average charge conductance, this time in units of the conductance quantum $e^2/h$, is given by,
\begin{equation}
\sum_{\sigma,\sigma'} |t_{i\sigma,j\sigma'}|^2
= \int_i {\rm d}y \int_j {\rm d}y_0 \sum_{\gamma,\gamma'}
\, A_\gamma A_{\gamma'} \exp[i (S_\gamma-S_{\gamma'})] {\rm Tr}[U_{\gamma'}^\dagger U_\gamma] \, ,
\end{equation}
and its average is usually calculated by performing the average
separately over orbital and spin degrees of freedom. 
Thus, in order to account for the spin-orbit 
effects, 
one multiplies
the right-hand side of Eq.~(\ref{eq:T})
by $\langle {\rm Tr}[U_{\gamma'}^\dagger U_\gamma] \rangle_{\rm SU(2)}$.
The leading-order contribution to the charge conductance is given by the
diagonal approximation, $\gamma=\gamma'$, with
 $\langle {\rm Tr}[U_{\gamma}^\dagger U_\gamma] \rangle_{\rm SU(2)} = {\rm Tr}[ I_{2 \times 2}] = 2$
 for spin 1/2 particles. The weak localization correction corresponds to the diagram shown in
 Fig.~\ref{fig:wloc}, which is multiplied by
 $\langle {\rm Tr}[U_{\rm loop}^2] \rangle_{\rm SU(2)} = -1$~\cite{Wal07}.
  If there is no spin rotation (in the absence
 of spin-orbit interaction), one instead obtains 2. This explains the magnitude and sign reversal of
 magnetoresistance with/without spin-orbit interaction.
 There are different ways to calculate
 such averages over SU(2). For instance one may write
 $$U_{\rm loop} = \left(
 \begin{array}{cc}
 \alpha & \beta \\
 -\beta^* & \alpha^*
 \end{array}
 \right) \, , $$
with $|\alpha|^2+|\beta|^2=1$, so that real and imaginary parts of $\alpha$ and $\beta$ correspond
to coordinates on a $3$-sphere. The average can then be calculated via an integral over the surface
of that sphere.

\begin{figure}[t]
 \centering
 \includegraphics[width=0.3\columnwidth]{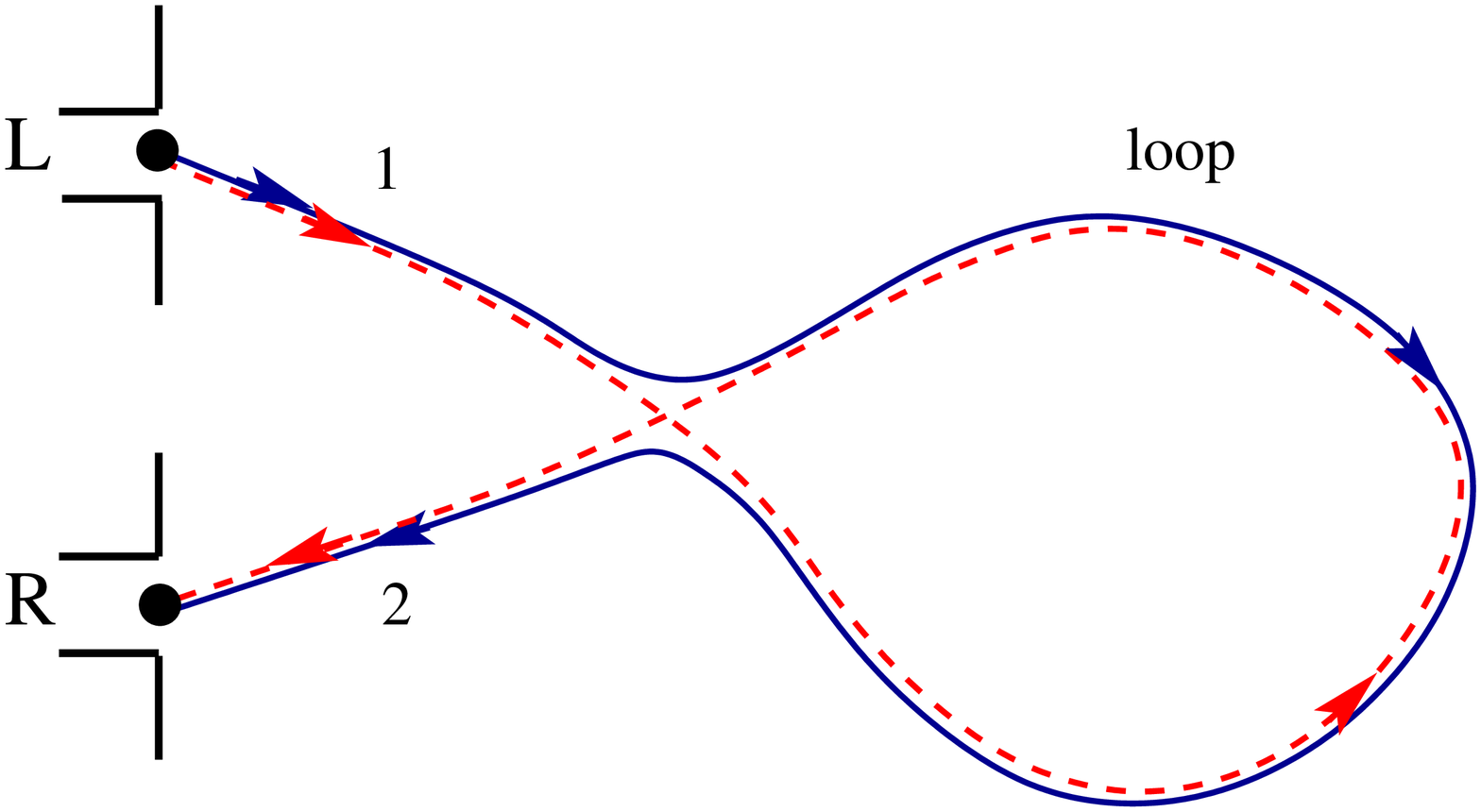}
\caption{\label{fig:wloc} (Color online) Semiclassical
diagram determining the weak localization contribution to the
conductance and spin conductance,
to leading order in the number $N \gg 1$ of transport channels. Blue (dark) and red
(light) trajectories travel in opposite direction along the loop, thus the contribution
vanishes in the presence of a large magnetic flux piercing the loop, giving
rise to magnetoresistance.\\[-7mm]}
\end{figure}

\begin{figure}[b]
 \centering
 \includegraphics[width=0.4\columnwidth]{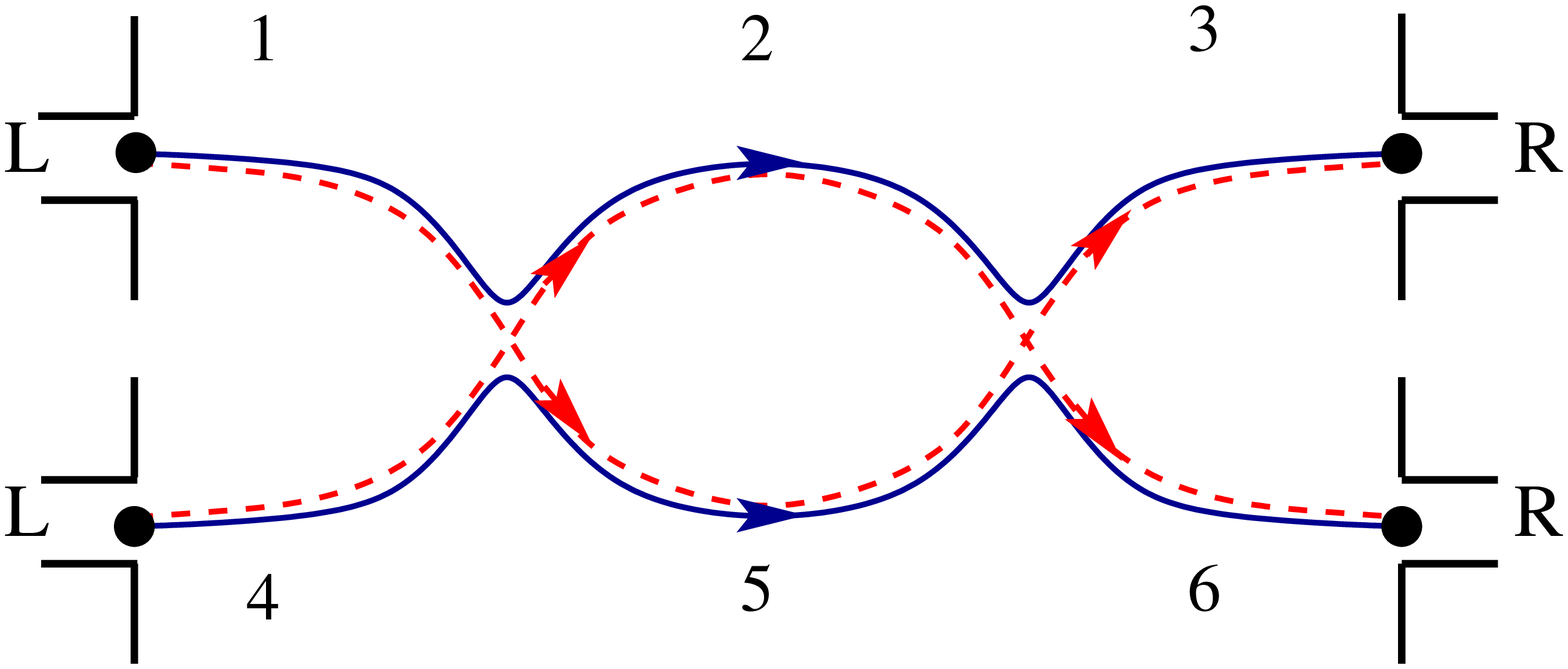}
\caption{\label{fig:ucf} (Color online) Contribution a) of Fig.1 of the main text
to the conductance fluctuations. Different trajectory segments are explicitly labelled.\\[-7mm]}
\end{figure}
For spin transport, on the other hand, the additional factor to calculate becomes
$\langle {\rm Tr}[U_{\gamma'}^\dagger \sigma^{(\mu)} U_\gamma] \rangle_{\rm SU(2)}$.
This average vanishes for both the diagonal and the weak localization contributions to the conductance~\cite{Ada10}.
For spin conductance fluctuations, the factors are obtained by labeling different trajectory segments.
This is done explicitly in Fig.~\ref{fig:ucf}. The spin-dependent prefactor is then straighforward to obtain,
for the spin conductance fluctuations it is
$\langle {\rm Tr}[U^\dagger_1 U^\dagger_5 U^\dagger_3 \sigma^{(\mu)} U_3 U_2 U_1] \times
{\rm Tr}[U^\dagger_4 U^\dagger_2 U^\dagger_6  \sigma^{(\mu)}  U_6 U_5 U_4] \rangle
= \langle {\rm Tr}[U^\dagger_5 U^\dagger_3 \sigma^{(\mu)} U_3 U_2] \times
{\rm Tr}[U^\dagger_2 U^\dagger_6  \sigma^{(\mu)}  U_6U_5] \rangle $. The calculation of this
average as an integral over the surface of the $3-$sphere does not present any technical difficulty, and
one finds that it vanishes. Similar labelling of the other contributions in Fig.1 of the main text lead to the
expression giving there, with only contributions d) and e) giving finite values as they are multiplied by
a different spin-prefactor.\\

{\bf The kicked rotator model for transport.}

Our numerics are based on the spin kicked rotator model.
The kicked rotator was introduced in the context of quantum chaos by Casati, Chirikov,  Izrailev and Ford
(for a review of the kicked rotator in quantum chaos see Ref.~\cite{Izr90}).
It is a generic model of dynamical systems. It has been extended to study open condensed matter
systems~\cite{Ric02,Two03}, where it has in particular been found that all properties
expected of ballistic quantum dots can be reproduced (weak localization, universal conductance fluctuations,
shot-noise and so forth). It has recently been extended to account for the presence
and the influence of spin-orbit interaction on charge transport in Ref.~\cite{Bar05}, again reproducing
expected reversal of magnetoresistance when the spin-orbit interaction is cranked up, the reduction in
conductance fluctuations and so forth. Ref.~\cite{Bar07} applied the spin kicked rotator to spin
transport, and it was found that the model reproduces random matrix theory predictions in a wide range of
parameters.

The Hamiltonian for the kicked rotator is
\begin{equation}
H = \frac{(k+l_0)^2}{2} + K \cos(p+\theta) \sum_n \delta(t-n \tau_0) \, ,
\end{equation}
which represents a free particle with kinetic energy $(k+l_0)^2/2$ periodically
perturbed by kicks of strength $K$ and period $\tau_0$. The latter time scale just serves
as a unit of time from now on and we accordingly set it equal to one. The parameters
$l_0$ and $\theta$ are necessary to break time-reversal symmetry~\cite{Izr90}. Because of
the system's additional symmetries two, and not one (e.g. magnetic field) parameters
are necessary to break time-reversal symmetry. The Hamiltonian is quantized on a torus
by discretizing momenta, $k \rightarrow k_l = 2 \pi l/M$, $l=1,2,... M$, and positions
$p \rightarrow p_n = 2 \pi n/M$. The model is usually represented by its Floquet, unitary
time-evolution operator from the middle of a free evolution period to the middle of the next one. In this way
the Floquet operator is symmetrized.
Momentum and position are related by a Fourier transform, so that writing the Floquet operator
in momentum representation requires two Fourier-transform sandwiched between the kick operator
$X = \exp[-i K \cos(p+\theta_0)]$ and the two half-period free-evolution operators. In this way one obtains Eq.(11) of the
main text, where $\Pi$ represents half-period free evolution matrices, $U$ give the Fourier transform
between momentum and position, and $X$ is the time-evolution operator corresponding to 
the kicked Hamiltonian. Spin physics may be introduced by extending the kick as in Eq.~(12) of the main text.

Finally transport through the system is introduced via two terminals in the form of two
absorbing strips in position coordinates. The $S$-matrix between these terminals is constructed
following Ref.~\cite{Two03}, leading to Eq.~(13) of the main text. 
For more details on the kicked rotator model, we refer the reader to 
Refs.~\cite{Ric02,Izr90,Two03,Bar05,Bar07}.

\end{widetext}

\end{document}